%% file: main.tex
\documentclass[sigconf]{acmart}
\usepackage{physics}
\usepackage{algorithm}
\usepackage[noend]{algpseudocode}
\usepackage{graphicx}
\usepackage{subcaption}
\usepackage{booktabs} 

\setcopyright{acmcopyright} 

\begin{document}
\title{Modelling Dynamic Interactions Between Relevance Dimensions}

\author{Sagar Uprety}
\affiliation{%
  \institution{The Open University}
  \streetaddress{Walton Hall}
  \city{Milton Keynes}
  \country{United Kingdom}
  \postcode{MK7 6AA}
}
\email{sagar.uprety@open.ac.uk}

\author{Shahram Dehdashti}
\affiliation{%
  \institution{Queensland University of Technology}
  \streetaddress{Gardens Point }
  \city{Brisbane}
  \country{Australia}
  \postcode{4000}
}
\email{shahram.dehdashti@qut.edu.au}

\author{Lauren Fell}
\affiliation{%
  \institution{Queensland University of Technology}
  \streetaddress{Gardens Point }
  \city{Brisbane}
  \country{Australia}
  \postcode{4000}
}
\email{l3.fell@qut.edu.au}

\author{Peter Bruza}
\affiliation{%
  \institution{Queensland University of Technology}
  \streetaddress{Gardens Point }
  \city{Brisbane}
  \country{Australia}
  \postcode{4000}
}
\email{p.bruza@qut.edu.au}

\author{Dawei Song}
\affiliation{%
  \institution{The Open University}
  \streetaddress{Walton Hall}
  \city{Milton Keynes}
  \country{United Kingdom}
} 

\additionalaffiliation{%
  \institution{Beijing Institute of Technology}
  \streetaddress{P.O. Box 1212}
  \city{Beijing}
  \country{China}
  \postcode{43017-6221}
}
\email{dawei.song@open.ac.uk}


\begin{abstract}
Relevance is an underlying concept in the field of Information Science and Retrieval. It is a cognitive notion consisting of several different criteria or dimensions. Theoretical models of relevance allude to interdependence between these dimensions, where their interaction and fusion leads to the final inference of relevance. We study the interaction between the relevance dimensions using the mathematical framework of Quantum Theory. It is considered a generalised framework to model decision making under uncertainty, involving multiple perspectives and influenced by context. Specifically, we conduct a user study by constructing the cognitive analogue of a famous experiment in Quantum Physics. The data is used to construct a complex-valued vector space model of the user's cognitive state, which is used to explain incompatibility and interference between relevance dimensions. The implications of our findings to inform the design of Information Retrieval systems are also discussed. 

\end{abstract}

\copyrightyear{2019} 
\acmYear{2019} 
\acmConference[ICTIR '19]{The 2019 ACM SIGIR International Conference on the Theory of Information Retrieval}{October 2--5, 2019}{Santa Clara, CA, USA}
\acmBooktitle{The 2019 ACM SIGIR International Conference on the Theory of Information Retrieval (ICTIR '19), October 2--5, 2019, Santa Clara, CA, USA}
\acmPrice{15.00}
\acmDOI{10.1145/3341981.3344233}
\acmISBN{978-1-4503-6881-0/19/10}

\maketitle
\input{introduction}

\input{related-work.tex}
\input{stern-gerlach.tex}
\input{cognitive-analogue.tex}
\input{exp-results.tex}

\input{results-discussion.tex}
\input{implication.tex}
\input{conclusion}
\bibliographystyle{ACM-Reference-Format}
\vspace{-3mm}
\input{acknowledgement.tex}
\vspace{-3mm}
\bibliography{bibliography}

\end{document}

%% file: introduction.tex
\vspace{-2mm}
\section{Introduction}
The concept of relevance lies at the heart of Information Retrieval (IR) and is fundamentally a cognitive notion, part of our cognitive ability. The underlying intent behind all the advances in IR has been to improve the relevance of information presented to the user. 

One of the main attributes of relevance is that it is a relation. There is always, implicitly or explicitly, the word `to' associated with relevance~\cite{Saracevic2016_notion}. It relates information or an information object to a context or situation. Relevance is also believed to manifest itself in different ways, with each manifestation indicating a different relation. Earlier works defined these manifestations at an abstract, philosophical level such as system relevance (related to the algorithmic query-document matching), topical relevance (related to subject expressed in the query), cognitive relevance (related to pertinence), situational relevance (related to utility), affective relevance (related to motivation/intent), etc. ~\cite{saracevic1997stratified, Saracevic2007_part2, COSIJN2000533_dimensions, cosijn2009relevance, Borlund2003TheIE}. In recent years, this plurality of relevance has been studied in terms of the judgement criteria considered by users. Apart from 'Topicality', there have been 'Reliability', 'Understandability', 'Novelty', 'Interest', etc. These relevance criteria are also called dimensions of relevance.

Another key attribute of relevance is that it is dynamic. It changes with user's interaction with information. In the Stratified model of relevance~\cite{saracevic1997stratified}, the different manifestations of relevance are considered as interacting layers. Each of these interacting strata include considerations or inferences of relevance. Our aim is to study the interaction between the different dimensions of relevance. In particular we study the effect of consideration of one relevance criterion on another. The main research question can be broken down into two sub-questions:

\textbf{\textit{RQ 1: How does consideration of one relevance criterion affect the inference of relevance with respect to another criterion? }}

\textbf{\textit{RQ 2: Can we construct a formal mathematical model of the user's underlying cognitive state in order to make predictions about such interactions?
}}

Going back to the stratified model of Saracevic~\cite{saracevic1996_stratified}, we can say that relevance inference takes place for each relevance dimension considered by the user and the final decision of relevance is a fusion of all the individual inferences. Our hypothesis is that\textit{ relevance inference at each dimension does not happen independently}, like a pre-defined value being read out of the internal cognitive state. It is rather constructed at the point of information interaction and thus influenced by the other dimensions considered by the user previously, which serve as a context for the inference of relevance for the current dimension. It is straightforward to see that consideration of an individual relevance criterion can affect the final judgement of relevance. For example, inferring that readability of a document is low may lead to a lower probability of judging it as relevant. Nevertheless, does consideration of readability as low also provide a context that affects the subsequent judgement about credibility of the document? 

We undertake a novel approach to test our hypothesis by constructing the cognitive analogue of the famous Stern-Gerlach experiment in Physics~\cite{fell:dehdashti:bruza:moreira:2019}. We conduct a user study where we show query-document pairs to participants and ask them questions about certain relevance dimensions in particular sequence. Through this process of elicitation, we construct a complex-valued Hilbert Space model of the underlying cognitive state. We ask the users questions about different relevance dimensions because not all users might consider the same relevance dimensions and in the same order while judging a document. Thus, it ensures consistency in our representation of the cognitive state for each user. The complex-valued Hilbert space model shows presence of quantum-like phenomena of incompatibility and interference in the decisions, and we verify the "quantumness" of our model using the Wigner function approach.

The main contributions of this paper are 1) a novel experimental design to construct a formal model which represents user's cognitive state for multidimensional relevance. It informs research in user behaviour, quantum-inspired IR and involves the use of non-commutating operators and complex numbers - something which has eluded quantum-inspired IR models in the past, and 2) Using the experiment design and the cognitive model, an investigation on the interaction between relevance dimensions which can inform IR system and user interface design.

In Section 2 we present some background about research in multidimensional relevance and Quantum Cognition. We then provide an easy-to-understand description of the Stern-Gerlach experiment in Section 3, followed by the construction of its cognitive analogue in Section 4. The details of our user study are given in Section 5, followed by an in-depth analysis of results in Section 6. We discuss the implications of our findings to IR in Section 7. Finally, Section 8 concludes the paper.

%% file: related-work.tex
\section{Background}
\subsection{Multidimensional Relevance}
Several works have investigated different factors, other than the query-document topical match, which users might consider in assessing relevance. One of the earliest works~\cite{cuadra_system1967experimental} investigating different relevant criteria identified 38 variables which effect relevant judgement. Later on, several studies were carried out in which users were asked to specify their judgement criteria~\cite{barry1994user_criteria, barry1998users_cross_sit,nilan_info_behav, nilan1988methodology, park1993nature_of_rel}. Scores of criteria such as depth/scope, accuracy, presentation quality, currency, tangibility, reliability, etc. were reported by users. In recent years, certain criteria or 'dimensions' are widely accepted as the most important considerations for user judgements of relevance. These include reliability~\cite{sm11_rel, yt11_rel, wnw14_rel, opla13_rel}, understandability~\cite{Understandability_Guido, Palotti_understandability}, novelty~\cite{Carbonell_mmr_diversity, zhai2003beyond_diversity}, effort~\cite{ja16_effort, vyc16_effort, yvcrb14_effort}, etc. A multidimensional relevance model was proposed ~\cite{ASI:Xu-MURM, MURM-psychometrics} which defined five such dimensions and was extended to seven dimensions in ~\cite{ASI:Jingfei}, including 'interest' and 'habit' dimensions. ~\cite{Jiang2017_in_situ} reported positive correlations between multiple relevance dimensions and user experience measures. Relevance judgement as an aggregate of the judgements under different dimensions was investigated in ~\cite{pasi_new_aggre_crit, daCostaPereira_pasi_priorit_agg_oper}. Dynamic fusion of relevance of different dimensions for ranking across sessions was proposed in ~\cite{Uprety:2018:MMU:3209978.3210130}. Thus we see that over the years research into multidimensional relevance has enriched our understanding of relevance and continues to inform search ranking and user behaviour understanding in IR.

\vspace{-3mm}
\subsection{Quantum Cognition}
In the  field of cognitive science, human cognitive states are approximated through the process of elicitation, with one assumption being that these internal states hold pre-defined values and elicitation acts to merely reveal such pre-existing values. However, research over the past few decades suggests that people's preferences or beliefs are often constructed during elicitation or judgement rather than read out from a pre-existing, definite state~\cite{Slovic1995_construc_pref}. The outcome of such a process of judgement also depends upon the context of judgement and the act of observation changes the internal state of belief. This has an analogue with the measurement of quantum systems which exist in an indefinite (superposition) state where the act of measurement creates a definite state. This analogy can be applied to certain fallacies in decision-making such as cognitive biases~\cite{Tversky1974} and order effects~\cite{Hogarth1992} which can be explained in terms of Quantum Interference and Incompatibility~\cite{Trueblood2011_quantum_account_ordereff, Pothos2009_quan_explan_irrational, Busemeyer2011_quantum_expl_prob_errors}. 
An Order effect is a phenomenon frequently encountered in human decision making where the different order of questions or evidences lead to different answers or decisions. Recently, order effects have been modelled using the mathematical framework of Quantum Theory~\cite{Trueblood2011_quantum_account_ordereff, Wang2013}. The fundamental advantage of using the quantum mathematical framework to model order effects lies in the use of a non-commutative algebra where events are represented by operators which do not necessarily commute with each other. 

The first study to consider interaction between relevance dimensions in terms of order effects was ~\cite{10.3389/fpsyg.2014.00612_Bruza}. In this work, the order of consideration of relevance criteria was found to manifest order effects. Thus, for example, the answer to the question about Reliability of a document was different when it was asked after the question of Understandability, than when asked first. Order effects in the presentation order of documents have been studied previously~\cite{Eisenberg1988_order, benyou_quantum_interf_Order, borlund_order, Xu2008_order, Huang2004_order}. The effect of order of relevance dimensions on clicked documents in query logs was investigated in ~\cite{Uprety:2018:IOE:3234944.3234972}.

%% file: stern-gerlach.tex
\vspace{-2mm}
\section{The Stern-Gerlach Experiment}
The Stern-Gerlach experiment (S-G) was one of the first experiments to show the necessity of a radical departure of modelling microscopic data from existing formalisms~\cite{sakurai}.

Consider a quantum system, say an electron. We focus on a particular property of an electron called the spin. A spin of an electron has two possible values - up or down (positive or negative). The spin is a magnetic property and it is possible to measure it by subjecting an electron beam towards the two poles of a magnet placed in a particular orientation. Those electrons which deflect towards the North pole of the magnet can be attributed, say, a positive spin and those who are deflected towards the opposite pole are said to have been in the negative spin state.

\begin{figure}[htb]
        \includegraphics[width=0.55\textwidth]{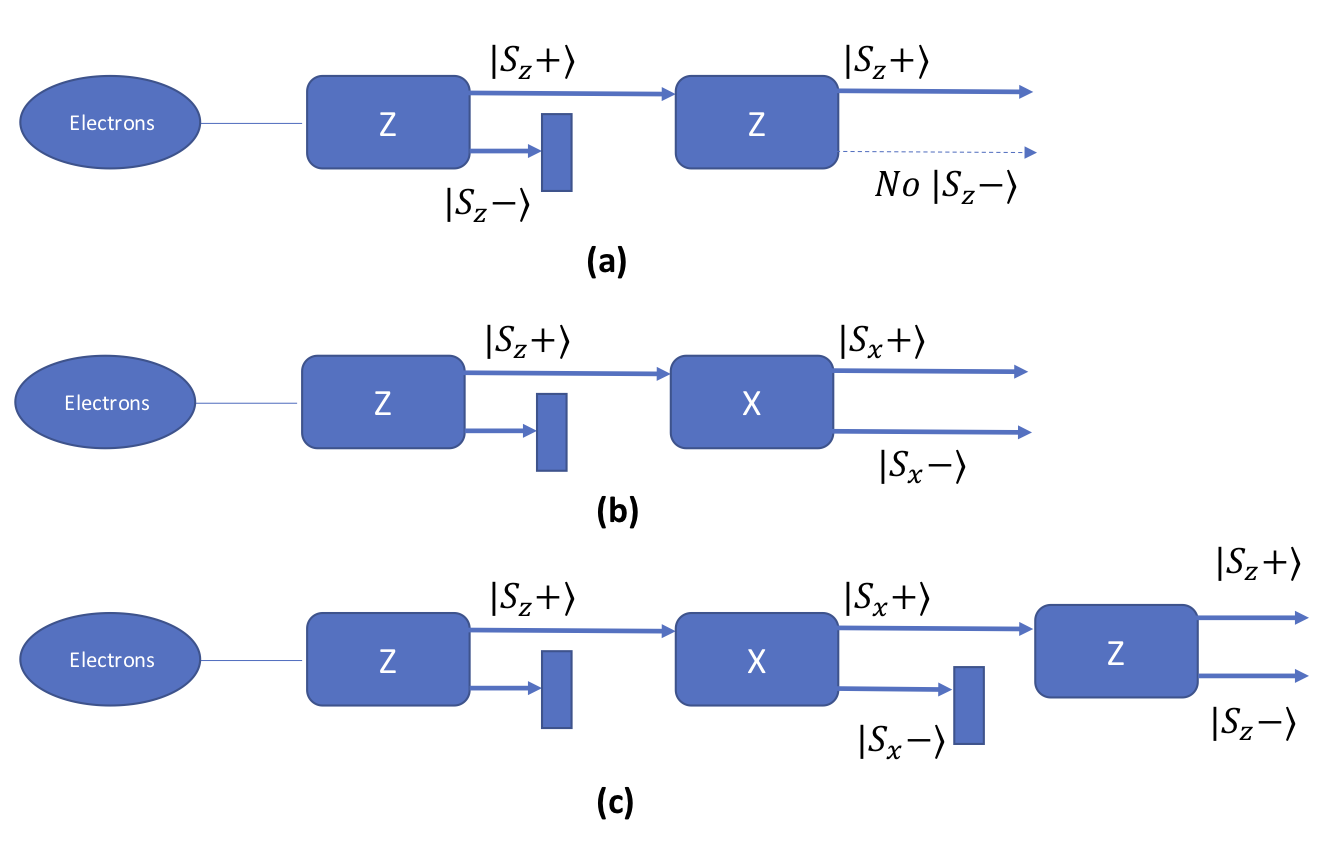}
        \caption{Stern-Gerlach Experiment}
        \label{figure-s-g}
        \vspace{-2mm}

\end{figure}

Now consider the series of experiments as shown in figure \ref{figure-s-g}. In the first setup (1.a), the negative spin electrons ($S_z^-$) coming out from the Z-axis apparatus are blocked and the spin positive electrons ($S_z^+$) are made to pass along the Z-axis apparatus once again. As expected, the output from the second Z-axis apparatus are all $S_z^+$ electrons. However, if instead of the second Z-axis apparatus, we put magnets along the X-axis, we find that half of the $S_z^+$ electrons deflect to the negative pole of the magnet ($S_x^-$) and half deflect towards the positive pole of the magnet kept along the X-axis ($S_x^+$). Thus, the positive or negative spin of an electron is not independent of the choice of measurement axis. Some of the electrons deflected towards the positive pole when measured along Z-axis are also getting deflected along the negative pole when measured along the X-axis. Things get weirder in setup \ref{figure-s-g}(c). A third Z- apparatus placed in the line of the $S_x^+$ electron beams shows presence of two beams - for the $S_z^+$ and $S_z^-$ spin states. This is despite that fact that $S_z^-$ was blocked after the first apparatus. It can be said that the measurement of $S_x^+$ component by the apparatus along the X-axis influences (in this case, completely destroys) any previous information about $S_z^+$ and $S_z^-$ ( i.e. the fact that we had all electrons in positive spin with respect to Z-axis and no $S_z^-$ components). 

In order to understand these results more clearly, we construct a model of these electron spins. We use the bra-ket notation to represent vectors. Any complex valued vector $A$ is represented as a ket - $\ket{A}$ and the complex conjugate of $A$ is a bra vector - $\bra{A}$. The inner product of two vectors $A$ and $B$ is calculated by taking the product of the bra of one vector and the ket of another - $\braket{B}{A}$. The norm of a vector is written as $|\braket{A}|^{1/2}$. As we saw, the $S_z^+$ electrons split equally into two directions when subject to magnets along X-axis, we represent the $S_z^+$ state of an electron as a linear combination of the states $S_x^+$ and $S_x^-$:
\vspace{-1mm}
\begin{equation} \label{s-z+}
    \ket{S_z^+} = \frac{1}{\sqrt{2}}\ket{S_x^+} + \frac{1}{\sqrt{2}}\ket{S_x^-}
\end{equation}
where the coefficients of state vectors $\ket{S_x^+}$ and $\ket{S_x^-}$ are called probability amplitudes and the square of these coefficients give the probability of finding an electron in a particular state. Here, an electron in $\ket{S_z^+}$ state is said to be in both $\ket{S_x^+}$ and $\ket{S_x^-}$ states at the same time, a concept called Superposition. Similar experiment with $S_z^-$ leads us to:
\vspace{-2.5mm}
\begin{equation} \label{s-z-}
    \ket{S_z^-} = \frac{1}{\sqrt{2}}\ket{S_x^+} - \frac{1}{\sqrt{2}}\ket{S_x^-}
\end{equation}
It is worth noting that the vectors $\ket{S_i^+}$ and $\ket{S_i^-}$ are orthogonal and therefore we have $\braket{S_i^+}{S_i^-} = 0$. Using this property, we can also express the X-axis spin states in terms of Z-axis spins:
\vspace{-1mm}
\begin{align} \label{s-x}
\ket{S_x^+} &= \frac{1}{\sqrt{2}}\ket{S_z^+} + \frac{1}{\sqrt{2}}\ket{S_z^-} \\ \nonumber
\ket{S_x^-} &= \frac{1}{\sqrt{2}}\ket{S_z^+} - \frac{1}{\sqrt{2}}\ket{S_z^-}
\end{align}
This explains our observation that the $S_x^+$ component from the second apparatus had both the $S_z^+$ and $S_z^-$ components. On careful examination of Equations \ref{s-z+},\ref{s-z-} and \ref{s-x}, we see that a definite state along Z-axis, say $\ket{S_z^+}$, is an indefinite state along the X-axis (as it is an equal superposition of both positive and negative spins). One cannot jointly determine both the Z and X component of the spin of the electron. These two properties are thus \textit{incompatible} with each other. 

To round up our explanation of the fundamentals of Quantum Theory through the S-G experiment, consider that instead of measuring along X-axis, we position the magnets along the Y-axis. We find a similar, symmetrical behaviour of electron spins such that we can consider the spin along the Y-axis to be in a superposition or linear combination of positive and negative spins along the Z-axis. So, we can represent the electron spin states along the Y-axis in terms of the states along Z-axis as: $\ket{S_y^+} = \frac{1}{\sqrt{2}}\ket{S_z^+} + \frac{1}{\sqrt{2}}\ket{S_z^-}$ and similarly for spin negative state along Y-axis. However, this makes $\ket{S_y^+} = \ket{S_x^+}$, but we know that the spin state of the electron along Y-axis exists separately as they get deflected onto the magnetic poles when aligned along the Y-axis. In order to resolve this issue, Quantum Theory turns towards complex numbers. We can represent the probability amplitudes of the states as complex numbers. Thus for the spin state along the Y-axis, we can write:
\vspace{-1mm}
\begin{align} \label{s-y}
\ket{S_y^+} &= \frac{1}{\sqrt{2}}\ket{S_z^+} + \frac{i}{\sqrt{2}}\ket{S_z^-} \\ \nonumber
\ket{S_y^-} &= \frac{1}{\sqrt{2}}\ket{S_z^+} - \frac{i}{\sqrt{2}}\ket{S_z^-}
\end{align}

Thus a two-dimensional vector space needed to describe the two-valued spin states of an electron along three different axis must be a complex vector space. 
\vspace{-3mm}

%% file: cognitive-analogue.tex
\vspace{-2mm}
\section{Cognitive Analogue of the S-G Experiment}
\subsection{Experiment Description in the Context of Relevance}
The cognitive analogue to the S-G experiment was originally discussed in ~\cite{fell:dehdashti:bruza:moreira:2019}.
In order to draw an analogy of the electron spin states in terms of human judgements, we consider the two-valued spin data to be equivalent to the yes/no answer data. The measurement along the different axes are equivalent to making judgements along different perspectives. So, for relevance judgements, one can consider positive spin and negative spin outcomes to be decisions of relevance and non-relevance respectively. The different axes are the different dimensions of considering relevance. Just like spin of an electron is not an independent quantity and depends on the axis of measurement, similarly relevance of a document cannot be assumed to exist independently of choice of dimension considered. A document may appear relevant when considering the topicality dimension but the users may be uncertain about its Reliability or Understandability. In our experiment, we consider three dimensions. Topicality - whether the information contained in the document is related to the topic of the query, Reliability - whether the user would rely on the information obtained from the document, and Understandability - how easy is it to understand the information presented in the document. We represent the cognitive state of a user before judging a document as:
\vspace{-2mm}
\begin{equation} \label{s-t}
    \ket{S} = t\ket{T+} + \sqrt{1-t^2}\ket{T-}
\end{equation}
 where $\ket{T+}$ represents the cognitive state of a user judging the document as topically relevant (with probability $t^2$) and $\ket{T-}$ represents the state of a user judging the document as topically irrelevant (with probability $1-t^2$). We could have represented the state $\ket{S}$ in terms of Reliability or Understandability states, but we choose the Topicality basis as the standard basis of representation. Similarly, we can represent the Understandability basis in terms of the Topicality basis as:
\vspace{-3mm}
\begin{align} \label{u-t}
    \ket{U+} &= u\ket{T+} + \sqrt{1-u^2}\ket{T-} \\ \nonumber
    \ket{U-} &= \sqrt{1-u^2}\ket{T+} - u\ket{T-}
\end{align}
Note that $\ket{U-}$ is constructed using the orthogonality constraint of $\ket{U+}$ and $\ket{U-}$. Here $u^2$ is the probability that users judge a document Understandable, given that they also consider it as Topical.

Further, in order to express Reliability dimension in terms of its interaction with the Topicality perspective, we write:
\vspace{-1.5mm}
\begin{align} \label{r-t}
    \ket{R+} &= r\ket{T+} + \sqrt{1-r^2}e^{i\theta_r}\ket{T-} \\ \nonumber
    \ket{R-} &= \sqrt{1-r^2}e^{-i\theta_r}\ket{T+} - r\ket{T-}
\end{align}
where, recall from Section 2, the need of using a complex probability amplitude for the third measurement basis. Thus, we need four parameters in order to construct the Hilbert space - $t, u, r$ and $\theta_r$. We intend to find these parameters by asking three sequential questions of each user analogous to performing measurements along different axes of the spin for a beam of electrons. 

For an initial state of the system $\ket{S}$, the probability of event $\ket{A}$ in the quantum framework is given by $P(A) = |\braket{A}{S}|^2$ i.e., square of projection of vector $\ket{S}$ onto vector $\ket{A}$. Note that the notation $\braket{A}{B}$ is the inner product of two vectors. The probability for event $A$ followed by $B$ is given as~\cite{Busemeyer:2012:QMC:2385442}:
\vspace{-2mm}
\begin{equation}\label{eqn-joint-prob}
P(B,A) = |\braket{B}{A}|^2|\braket{A}{S}|^2
\end{equation}
which is read from right to left as projecting the initial state $\ket{S}$ to the vector for event $A$ and then projecting this state (to which the initial state has collapsed) onto the state vector for event $B$.
The quantum framework does not define joint probability of events $A$ and $B$ as, in general, $P(A,B) \neq P(B,A)$. As we can see $P(A,B) = |\braket{A}{B}|^2|\braket{B}{S}|^2$, which for $\braket{A}{S} \neq \braket{B}{S}$ is not equal to $P(B,A)$ in Equation \ref{eqn-joint-prob}. Note that we use the notation $P(B,A)$ to refer to the probability of the sequence $A->B$, i.e. $B$ takes place after event $A$.
\vspace*{-5mm}
\subsection{Experimental Design}

Our experiment thus consists of the following steps:

\begin{enumerate}
    \item We first prepare a user's cognitive state into one of $\ket{T+}$ or $\ket{T-}$ states by asking a user whether a document is topically related to the query or not. This consists of projecting the user's cognitive state $\ket{S}$ onto the vectors $\ket{T+}$ and $\ket{T-}$. 
    Thus the probability of obtaining a positive response on asking the question about topicality from a user is given as $P(T+) = |\braket{T+}{S}|^2 = t^2$. We can thus obtain the value for $t$.
    \item
    \begin{enumerate}
        \item Next, we take the users who answered yes to the topicality question and ask them about the understandability of the document. We can obtain the probability $P(U+, T+)$. This is represented in the Hilbert space as
        \begin{equation}
            P(U+,T+) = |\braket{U+}{T+}|^2|\braket{T+}{S}|^2 = u^2*t^2
        \end{equation}
        (Note from Equation \ref{u-t} that $\braket{U+}{T+} = u$). We thus obtain value of $u$.
        
        \item Instead of asking the question about understandability, if we take some of the users who respond positively to topicality and ask them about reliability of the document, we can calculate the probability \\
        $P(R+,T+) = |\braket{R+}{T+}|^2|\braket{T+}{S}|^2 = r^2*t^2$ and thus obtain the value of $r$.
    \end{enumerate} 
    
    \item Now we are left with figuring out the value of $\theta_r$. This is done in the following way - those users who answer positively to topicality and understandability questions are asked the third question about reliability. Thus 
    \begin{equation}
        P(R+,U+,T+) =|\braket{R+}{U+}|^2|\braket{U+}{T+}|^2|\braket{T+}{S}|^2
    \end{equation}
    Note that $\braket{R+}{U+}$ is a complex quantity and its square is calculated by multiplying it by its complex conjugate. Thus $|\braket{R+}{U+}|^2 = \braket{U+}{R+}\braket{R+}{U+}$. Hence we have, 
    \begin{align}
        \braket{U+}{R+} &= (u\bra{T+} + \sqrt{1 - u^2}\bra{T-})\times (r\ket{T+} + \sqrt{1-r^2}e^{i\theta_r}\ket{T-}) \\ \nonumber
        &= ur + \sqrt{(1-u^2)(1-r^2)}e^{i\theta_r} \\ \nonumber
        \braket{R+}{U+}&=  |\braket{U+}{R+}|^{+} \\ \nonumber
        &= ur + \sqrt{(1-u^2)(1-r^2)}e^{-i\theta_r}
    \end{align}
    Finally, 
    \begin{equation}
    \braket{U+}{R+}\braket{R+}{U+} = (ur)^2 + (1-u^2)(1-r^2) + 2ur\sqrt{(1-u^2)(1-r^2)}\cos{\theta_r}
    \end{equation}
    Now, we know $u$ and $r$ from previous steps, the probability $ P(R+,U+,T+)$ obtained from the experimental data helps us to calculate the value of $\theta_r$
 
\end{enumerate}

%% file: exp-results.tex
\vspace{-2mm}
\section{Experiment}
\vspace{-1mm}
\subsection{Participants}
We recruited 300 participants for a user study using the online crowd-sourcing platform Prolific (prolific.ac). The only pre-screening criterion for participating in the study was a cut-off of 96 percent approval rate. Approval rate for a participant in Prolific is the fraction of submitted responses approved. The participants were paid at a rate of \textsterling7.08 per hour. Data of 5 participants was excluded as they completed the study in much less time than the minimum duration assumed for proper responses. The questionnaire was designed using the Qualtrics platform (qualtrics.com/uk). Proper consent was sought and they were also informed that data protections laws are being complied with. The study was approved by The Open University UK's OU Human Research Ethics Committee with reference number HREC/3063/Uprety.
\vspace{-2mm}
\subsection{Material}
\begin{table*}[htb]
\centering
\resizebox{\textwidth}{!}{
 \begin{tabular}{|c|c|c|} 
 \hline
Query & Information Need & Source \\
\hline
Radio Waves and Brain Cancer & Look for evidence that radio waves from radio towers or mobile phones affect brain cancer occurrence & TREC 2005 Robust track (310) \\
\hline
symptoms of mad cow disease in humans & Find information about mad cow disease symptoms in humans  & TREC 2013 Web Track (236) \\
\hline
educational advantages of social networking sites & What are the educational benefits of social networking sites? & TREC 2014 Web Track (293) \\
\hline
 \end{tabular}
 }
\caption{Selected queries and their descriptions}
\label{table-query_descriptions}
\vspace{-6mm}
\end{table*}
The participants were shown three queries and one document snippet for each query as it appears in popular search engines like Google and Bing. The queries and description of the information need (IN) were shown as consistent with the TREC style, as listed in Table \ref{table-query_descriptions}. The document snippets were constructed manually by altering particular aspects of existing documents obtained in order to introduce both uncertainty in judging with respect to a particular dimension and also incompatibility between the dimensions.

\begin{figure}[htb]
        \includegraphics[width=0.50\textwidth]{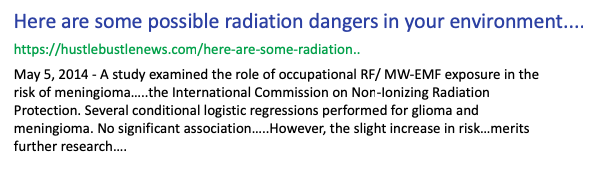}
        \caption{Snapshot of a document}
         \vspace{-4mm}
        \label{figure-study_snapshot}
\end{figure}

For instance, Figure \ref{figure-study_snapshot} shows the snippet for the first query. The source URL is created in a way as to create some uncertainty about the reliability of the source. In the same way, the title of the document does not explicitly reflect that it is about the topic of the query and it is also not easy for everyone to understand the information in the body of the snippet. An uncertain user might answer negatively to the topicality question (attains the definite $\ket{T-}$ state), but on being asked to consider the understandability dimension, the user might read the snippet body carefully which can influence the user to become uncertain about topicality again. Similar criteria was followed in designing the document snippets for the other two queries, which could not be shown here due to limited space. The three queries chosen thus allowed us to design document snippets to exhibit these characteristics. 

\begin{table}[h!]
\centering
\begin{tabular}{|c|c|c|c|} 
 \hline
\textbf{Parameter}& \textbf{Query 1} & \textbf{Query 2} &\textbf{ Query 3 }\\
\hline
$P(T+)$ & $0.7622$ & $0.6736$ & $0.8993$ \\
\hline
$P(U+,T+)$ & $0.4405$ & $0.5416$ & $0.8724$ \\
\hline
$P(R+,T+)$ & $0.4609$ & $0.4857$ &  $0.5616$ \\
\hline
$P(R+,U+,T+)$ & $0.2587$ & $0.4513$ & $0.6442$ \\
\hline
$P(R+,U-,T+)$ & $0.1188$ & $0.0694$ & $0.0000$ \\
\hline
$P(U+,R+,T+)$ & $0.2765$ & $0.4285$ & $0.5410$ \\
\hline
$P(U+,R-,T+)$ & $0.1560$ & $0.0857$ & $0.2739$ \\
\hline
$t^2$ & $0.7622$ & $0.6736$ & $0.8993$ \\
\hline
$u^2$ & $0.5779$ & $0.8041$ & $0.9701$ \\
\hline
$r^2$ & $0.5462$ & $0.7311$ & $0.6456$ \\
\hline
$\theta_r$ & $80.62 \deg$ &$56.79 \deg$ & $51.43 \deg$\\
\hline
\end{tabular}
\caption{Parameter values and associated probabilities}
\label{table-results}
\vspace{-8mm}
\end{table}
\subsection{Procedure}
The participants were shown the query and the document and after that asked the following questions:

\begin{enumerate}
    \item Is the document about the topic of the search query? (T)
    \item Is it easy to understand the information presented in the document snippet? (U)
    \item Would you rely on the information presented in this document? (R)
\end{enumerate}
Note that a between subjects design was carried out and the participants were uniformly split into two groups - one group asked questions in the TUR sequence, which is used to calculate parameters $u$ and $\theta_r$, by calculating probabilities $P(U+,T+)$ and $P(R+,U+,T+)$. The other group asked questions in the TRU sequence, in order to calculate the parameter $r$, by calculating $P(R+,T+)$. The participants were shown the next question only after answering the current question, so that they their answers are not primed by seeing all the three questions.

%% file: results-discussion.tex
\vspace{-2mm}
\section{Results and Discussion}
Some of the probabilities obtained and the required parameter values for constructing the Hilbert space are shown in Table \ref{table-results}. As we see, we get the complete two-dimensional Hilbert space involving the 3 real parameters and the complex phase $\theta_r$. 3 questions are necessary because it implies measurement along 3 different basis which gives rise to the need for complex number representation. The existence of a superposition state signifies that initially a user does not exist in a definite state of judgement with respect to an information object. When a particular question is asked or is considered in the user's mind, e.g. about reliability, this uncertainty resolves and the user's cognitive state collapses to one of two values of relevance or non-relevance.
\vspace{-3mm}
\subsection{Wigner Function}
We constructed a complex-valued Hilbert space to model the user cognitive state for decisions from incompatible perspectives. In doing so, we utilised the mathematical framework of quantum theory. An important question to ask is why do we need quantum theory to model such a decision-making scenario? We verify the "quantumness" of the model using a criterion in quantum theory which distinguishes between quantum and classical statistics in the data. Quantum theory has a range of such criteria, one of which is the discrete Wigner function~\cite{Wootters1987_wigner, Galvo2005_wigner}. 
Wigner functions are quasi-probability distributions which map quantum states to a phase space. Quasi-probability distributions relax some of the axioms of the Kolmogorov probability theory, the highlight being the existence of negative probabilities in the distribution for states which do not have a classical model. The discrete Wigner function distribution is given as:
\vspace{-2mm}
\begin{equation}
    W = \begin{pmatrix} 1 + r_x + r_z & 1 - r_x + r_z \\ 1 - r_x - r_z & 1 + r_x - r_z \end{pmatrix}
\end{equation}
where $r_x = 2*\sqrt{t^2(1 - t^2)}$ and $r_z = 2*t^2 - 1$.
We omit the derivation due to lack of space but the reader can refer to \cite{fell:dehdashti:bruza:moreira:2019}.
The Wigner function for the three query-document pairs is obtained as:
\vspace{-1.5mm}
\begin{align}
    W_1 &= \begin{pmatrix} 0.5939 & 0.1683 \\ -0.0939 & 0.3317 \end{pmatrix}
    \hspace{2mm} W_2 = \begin{pmatrix} 0.5712 & 0.1024 \\ -0.0712 & 0.3976 \end{pmatrix} \\ \nonumber
    W_3 &= \begin{pmatrix} 0.6001 & 0.2992 \\ -0.1001 & 0.2008 \end{pmatrix}
\end{align}
The negative values in the Wigner function distribution is an indicator of quantum interference which shows that the statistics generated by our Stern-Gerlach type experiment are quantum statistics and thus a quantum model is needed to model such data. As discussed before, the interference effects are due to the incompatibility between decision perspectives. The decision of reliability interferes with that of understandability, for example. 
\vspace{-3mm}
\subsection{Incompatibility}
In Quantum Theory, incompatibility of measurements can be represented in the form of non-commutating operators. Operators are matrices which encapsulate a measurement which can be performed on a quantum state. Measuring a property of a system generates an event. We can construct operators by first constructing their eigenvectors, also called projectors. The projector of an event $A$ is represented by the outer product of the vector corresponding to $A$ with itself, i.e $\ket{A}\bra{A}$. In the complex-valued two-dimensional Hilbert space we constructed, we have 3 basis, corresponding to $T, R$ and $U$ questions/measurements. We assume $Topicality$ as the standard basis, and hence the orthogonal vectors $\ket{T+}$ and $\ket{T-}$ are given as:
\vspace{-2mm}
\begin{equation} \label{eqn_standard_basis}
     \ket{T+} = \begin{pmatrix} 1 \\ 0 \end{pmatrix}
     \hspace{2mm}
     \ket{T-} = \begin{pmatrix} 0 \\ 1 \end{pmatrix}
\end{equation}
Thus the projector for event $T$ is given by:
\begin{align}
    \ket{T+}\bra{T+} = \begin{pmatrix} 1 & 0 \\ 0 & 0 \end{pmatrix}
    \hspace{3mm}
     \ket{T-}\bra{T-} = \begin{pmatrix} 0 & 0 \\ 0 & 1 \end{pmatrix}
\end{align}
The two projectors form the eigen vectors of the operator for the event $T$ with eigen values $+1$ and $-1$. Thus we have the operator $T$ as :
\vspace{-4mm}
\begin{align} \label{eqn_T_operator}
    \hat{T} &= \ket{T+}\bra{T+} - \ket{T-}\bra{T-} \\ \nonumber
    &= \begin{pmatrix} 1 & 0 \\ 0 & -1 \end{pmatrix}
\end{align}

Combining Equations \ref{eqn_standard_basis} and \ref{u-t}, we can write the vectors for event $U$ as:
\vspace{-4mm}
\begin{align}
         \ket{U+} = \begin{pmatrix} 0.7601 \\ 0.6496 \end{pmatrix}
     \hspace{2mm}
     \ket{U-} = \begin{pmatrix} 0.6496 \\ -0.7601 \end{pmatrix}
\end{align}

Thus we get the operators for $U$ and $R$ for query $1$ as:
\vspace{-1mm}
\begin{align} \label{eqn_U_operator}
    \hat{U} &= \begin{pmatrix} 0.1558 & 0.9874 \\ 0.9874 & -0.1558 \end{pmatrix}
    \hspace{2mm}
    \hat{R} &= \begin{pmatrix} 0.0924 & 0.9955e^{i80.62} \\ 0.9955e^{-i80.62} & -0.0924 \end{pmatrix}
\end{align}
We find that all the three operators do not commute pairwise - $[\hat{T},\hat{U}] \neq 0, [\hat{T},\hat{R}] \neq 0, [\hat{R},\hat{U}] \neq 0$ for the three queries. The consequence of incompatibility is that it is not possible to form joint distributions over answers involving incompatible questions. Thus $P(T=+, R=+)$ is not defined because it will be different if order of questions are different, i.e., $P(T=+, R=+) \neq P(R=+, T=+)$. 

This is where the advantage of using the Quantum framework lies. In classical probability, events always commute and thus order effects cannot be modelled. Order effects are bound to occur when the relevance dimensions are considered in different orders. From a cognitive point of view, a user is unable to be in a certain state of decision using two relevance criteria. Certainty in one relevance criterion does not imply certainty in another incompatible criterion. The method of constructing incompatible operators formally establishes and predicts order effects.
\vspace{-3mm}
\subsection{Interference between Dimensions}
Our main research question is to investigate the effect of consideration of one dimension on the judgement with respect to another dimension. More specifically, in our experiment, we are testing the interaction between Understandability and Reliability. We ask all users the question of Topicality first because Topicality is generally the foremost criterion of judging a document. We suspect that judgement of Reliability will be effected by whether or not Understandability has been considered. By consideration of Understandability, it is meant that the user has made an effort in comprehending the content of the document. 

In Table \ref{table-R-given-U}, we compare the probability of answering 'Yes' to Reliability question after the Topicality question, with the probability obtained had Understandability been answered before. We see that when users are unable to understand the document, they do not find it reliable either. Statistically significant results are reported for queries 1 and 2, shown in bold font in the table (Chi-square two tailed test of the equality of proportions, $\alpha = 0.05$). On the other hand, although we see that having comprehended the information better increases the probability of judging it more Reliable, the increase in probability is not statistically significant.

We also see that Reliability has a similar effect on Understandability as shown in Table \ref{table-U-given-R}. Those users who do not find the documents Reliable don't find it Understandable either. Intuitively one feels that Understandability should be independent of the Reliability of the document, but the data shows the dependence. We hypothesise that users who do not find the document Reliable do not make much effort to judge the Understandability dimension and hence the high correlation. 

Interference is another implication of incompatibility in decision making which is witnessed in decision data as Law of Total Probability (LTP) violation. For the participants who have answered the question about 'Understandability' first and then 'Reliability', we calculate the probability of answering 'Yes' to 'Reliability' using the law of total probability (LTP) as:
\vspace{-1mm}
\begin{equation}\label{LTP}
P_u(R+,T+) = P(R+,U+,T+) + P(R+,U-,T+)
\end{equation}
The probabilities on the two sides of the above equation are calculated from the data and reported in Table \ref{table-ltp}. We see that $P_u(R+) \neq P(R+)$. However, none of the results are statistically significant. We suspect that even when users are judging Reliability without being asked about Understandability, some of them do consider it in their mind (also due to learning from judging the first query). This is equivalent to being asked the question about Understandability as a form of self-elicitation which creates a definite belief state with respect to Understandability. Therefore we do not see a statistically significant difference in the probabilities in the two situations. As such, it is difficult to segregate judgements made by only considering Reliability, from those considering Understandability before Reliability. However, when they do consider Understandability before Reliability, it does make a difference in judgement of Reliability (as discussed in the above two paragraphs).

Note that the calculation of $P_u(R+,T+)$ in the quantum framework incorporates the interference term, which is a function of the complex phase $\theta_r$, which is able to model this interference in experimental data.
\vspace{-1mm}
\begin{equation}
P_u(R+,T+) = P(R+,U+,T+) + P(R+,U-,T+) + Int(\theta_r)
\end{equation}

\begin{table}[h!]
\vspace{-3mm}
\centering
{\begin{tabular}{|c|c|c|c|} 
 \hline
& \textbf{Q1} & \textbf{Q2} & \textbf{Q3}\\ 
\hline
P(R+|T+) & 0.5462 & 0.7311 & 0.6456 \\
\hline
P(R+|U+,T+) & 0.5872 & 0.8332 & 0.7384\\
\hline
P(R+|U-,T+) & \textbf{0.3692} & 0.5261 & \textbf{0.0000}\\
\hline
\end{tabular}
\caption{Effect of Understandability on Reliability}
\label{table-R-given-U}
}
\vspace{-6mm}
\end{table}

\begin{table}[h!]
\centering
{\begin{tabular}{|c|c|c|c|} 
 \hline
& \textbf{Q1} & \textbf{Q2} & \textbf{Q3}\\ 
\hline
P(U+|T+) & 0.5779 & 0.8040 & 0.9701 \\
\hline
P(U+|R+,T+) & 0.5999 & 0.8822 & 0.9633\\
\hline
P(U+|R-,T+) & \textbf{0.4074} & \textbf{0.4801} & 0.8887\\
\hline
\end{tabular}
\caption{Effect of Reliability on Understandability }
\label{table-U-given-R}
}
\vspace{-6mm}
\end{table}

\begin{table}[h!]
\centering
{\begin{tabular}{|c|c|c|} 
 \hline
\textbf{Query} & \textbf{$P(R+,T+)$} & \textbf{$P_u(R+,T+)$} \\ 
\hline
Query 1 & 0.3775 & 0.4609  \\
\hline
Query 2 & 0.5207 & 0.4857\\
\hline
Query 3 & 0.6442 & 0.5616 \\
\hline
\end{tabular}
\caption{Interference as violation of LTP}
\label{table-ltp}
}
\vspace{-6mm}
\end{table}

%% file: implication.tex
\vspace{-4mm}
\section{Implications to IR System Design}
This study and previous studies on order effects in IR show 1) Consideration of a particular relevance dimension has an effect on judgement of a subsequent dimension, 2) The order of consideration of these dimensions effects the final judgement. The role of an IR system is to provide the user with relevant information which helps the user accomplish some task or make a well-informed decision. Therefore, it is important that users are able to reconcile the different dimensions of relevance in a way which enables them to select the most relevant information for their need. 

There are a few different ways in which IR systems can either circumvent or exploit the interference and incompatibility effects, accordingly so as to maximise the probability of the user finding relevant information. Firstly, IR systems can help reduce uncertainty in judging information objects by providing the users extra information about different relevance dimensions. For example, a news retrieval system can provide, along with each article, a score for dimensions like credibility, readability (understandability), factuality, opinionated-ness, etc. These scores of Information Nutrition~\cite{Fuhr_info_nutrition} can be calculated based on information object content, or/and be collected through user provided data. This can help users reduce uncertainty in judging the information or information object, and, more importantly make them consider the optimised sequence of dimension in order to make the best possible judgement.

Secondly, the order of consideration of relevance dimensions by a user can be ascertained and documents be ranked in that order. For example, for the query 'Game of Thrones news', a popular Television series, a cautious user might look for highly reliable articles but a more adventurous user might consider articles which appear 'Interesting', talking about different conspiracy theories or spoilers about the TV show, the credibility of such articles being a secondary criterion. The IR system can present documents according to the dimensional preference of the user if it is able to profile the user appropriately. On other hand, the order of preference of relevance criteria might be largely independent of the user but depend upon the type of information need, e.g., for 'Visa to US' queries, users may always prefer reliability of the source as the first criterion to judge (along with topicality, of course). A related approach has been proposed in \cite{pasi_new_aggre_crit,daCostaPereira_pasi_priorit_agg_oper} where, taking inspiration from the Multi-Criteria Decision Making (MCDM) approach in the area of Decision Theory, a prioritised aggregation operator is defined for different types of users. Our work informs this approach by exploring the dynamic interaction between the relevance dimensions themselves.

Thirdly, this work would also inform the design of relevance assessment collection procedure. The annotators could be given a list of relevance dimensions and asked to rate the relevance of a document along each of these dimensions, apart from a final relevance judgement. This will help us capture more context surrounding the relevance assessment of a document. Also, instead of having a fixed list, they can choose the order of dimensions. This method can reveal whether users prefer a particular order and whether having a fixed order or a random order of dimensions considered for relevance judgement lead to the same or different final judgement. 

Lastly, the construction of the Hilbert space-based user model offers a principled mathematical modelling approach to user studies in IR. Not only do we investigate user behaviour in multidimensional judgement, we parallelly construct a vector space model in order to be able to predict such behaviour. 

%% file: conclusion.tex
\vspace{-3.5mm}
\section{Conclusion and Future Work}
In this study we have attempted to investigate the interaction between select relevance dimensions in the form of context effects. The experiment is designed in a way so as to capture these context effects. A formal model using a complex-valued Hilbert space for the user's cognitive state is constructed. Our hypothesis that relevance dimensions effect each other is shown by the presence of incompatibility and interference/context effects. To our knowledge, this is also the first work in quantum-inspired IR where complex numbers arise naturally in decision making. For our future work, we intend to investigate different forms of interactions between relevance dimensions and work towards building fusion models for multidimensional relevance by making use of the complex Hilbert space model. This will help in building more accurate user models for decision fusion under incompatibility and also inform the users to make best possible judgement.

%% file: acknowledgement.tex
\section*{Acknowledgements}
Authors affiliated to Open University, UK are funded by the European Union's Horizon 2020 research and innovation programme under the Marie Sklodowska-Curie grant agreement No 721321. Authors affiliated to Queensland University of Technology, Australia are supported by the Asian Office of Aerospace Research and Development (AOARD) grant: FA2386-17-1-4016.

%% file: main.bbl

\begin{thebibliography}{00}


\ifx \showCODEN    \undefined \def \showCODEN     #1{\unskip}     \fi
\ifx \showDOI      \undefined \def \showDOI       #1{#1}\fi
\ifx \showISBNx    \undefined \def \showISBNx     #1{\unskip}     \fi
\ifx \showISBNxiii \undefined \def \showISBNxiii  #1{\unskip}     \fi
\ifx \showISSN     \undefined \def \showISSN      #1{\unskip}     \fi
\ifx \showLCCN     \undefined \def \showLCCN      #1{\unskip}     \fi
\ifx \shownote     \undefined \def \shownote      #1{#1}          \fi
\ifx \showarticletitle \undefined \def \showarticletitle #1{#1}   \fi
\ifx \showURL      \undefined \def \showURL       {\relax}        \fi
\providecommand\bibfield[2]{#2}
\providecommand\bibinfo[2]{#2}
\providecommand\natexlab[1]{#1}
\providecommand\showeprint[2][]{arXiv:#2}

\bibitem[\protect\citeauthoryear{Barry}{Barry}{1994}]%
        {barry1994user_criteria}
\bibfield{author}{\bibinfo{person}{Carol~L Barry}.}
  \bibinfo{year}{1994}\natexlab{}.
\newblock \showarticletitle{User-defined relevance criteria: An exploratory
  study}.
\newblock \bibinfo{journal}{{\em Journal of the American Society for
  Information Science\/}} \bibinfo{volume}{45}, \bibinfo{number}{3}
  (\bibinfo{year}{1994}), \bibinfo{pages}{149--159}.
\newblock


\bibitem[\protect\citeauthoryear{Barry and Schamber}{Barry and
  Schamber}{1998}]%
        {barry1998users_cross_sit}
\bibfield{author}{\bibinfo{person}{Carol~L Barry} {and} \bibinfo{person}{Linda
  Schamber}.} \bibinfo{year}{1998}\natexlab{}.
\newblock \showarticletitle{Users' criteria for relevance evaluation: a
  cross-situational comparison}.
\newblock \bibinfo{journal}{{\em Information processing \& management\/}}
  \bibinfo{volume}{34}, \bibinfo{number}{2-3} (\bibinfo{year}{1998}),
  \bibinfo{pages}{219--236}.
\newblock


\bibitem[\protect\citeauthoryear{Borlund}{Borlund}{2003}]%
        {Borlund2003TheIE}
\bibfield{author}{\bibinfo{person}{Pia Borlund}.}
  \bibinfo{year}{2003}\natexlab{}.
\newblock \showarticletitle{The IIR evaluation model: a framework for
  evaluation of interactive information retrieval systems}.
\newblock \bibinfo{journal}{{\em Inf. Res.\/}}  \bibinfo{volume}{8}
  (\bibinfo{year}{2003}).
\newblock


\bibitem[\protect\citeauthoryear{Bruza and Chang}{Bruza and Chang}{2014}]%
        {10.3389/fpsyg.2014.00612_Bruza}
\bibfield{author}{\bibinfo{person}{Peter Bruza} {and} \bibinfo{person}{Vivien
  Chang}.} \bibinfo{year}{2014}\natexlab{}.
\newblock \showarticletitle{Perceptions of document relevance}.
\newblock \bibinfo{journal}{{\em Frontiers in Psychology\/}}
  \bibinfo{volume}{5} (\bibinfo{year}{2014}), \bibinfo{pages}{612}.
\newblock
\showISSN{1664-1078}
\showDOI{%
\url{https://doi.org/10.3389/fpsyg.2014.00612}}


\bibitem[\protect\citeauthoryear{Busemeyer and Bruza}{Busemeyer and
  Bruza}{2012}]%
        {Busemeyer:2012:QMC:2385442}
\bibfield{author}{\bibinfo{person}{Jerome~R. Busemeyer} {and}
  \bibinfo{person}{Peter~D. Bruza}.} \bibinfo{year}{2012}\natexlab{}.
\newblock \bibinfo{booktitle}{{\em Quantum Models of Cognition and Decision\/}
  (\bibinfo{edition}{1st} ed.)}.
\newblock \bibinfo{publisher}{Cambridge University Press},
  \bibinfo{address}{New York, NY, USA}.
\newblock
\showISBNx{110701199X, 9781107011991}


\bibitem[\protect\citeauthoryear{Busemeyer, Pothos, Franco, and
  Trueblood}{Busemeyer et~al\mbox{.}}{2011}]%
        {Busemeyer2011_quantum_expl_prob_errors}
\bibfield{author}{\bibinfo{person}{Jerome~R. Busemeyer},
  \bibinfo{person}{Emmanuel~M. Pothos}, \bibinfo{person}{Riccardo Franco},
  {and} \bibinfo{person}{Jennifer~S. Trueblood}.}
  \bibinfo{year}{2011}\natexlab{}.
\newblock \showarticletitle{A quantum theoretical explanation for probability
  judgment errors.}
\newblock \bibinfo{journal}{{\em Psychological Review\/}}
  \bibinfo{volume}{118}, \bibinfo{number}{2} (\bibinfo{year}{2011}),
  \bibinfo{pages}{193--218}.
\newblock
\showDOI{%
\url{https://doi.org/10.1037/a0022542}}


\bibitem[\protect\citeauthoryear{Carbonell and Goldstein}{Carbonell and
  Goldstein}{1998}]%
        {Carbonell_mmr_diversity}
\bibfield{author}{\bibinfo{person}{Jaime Carbonell} {and} \bibinfo{person}{Jade
  Goldstein}.} \bibinfo{year}{1998}\natexlab{}.
\newblock \showarticletitle{The Use of MMR, Diversity-based Reranking for
  Reordering Documents and Producing Summaries}. In \bibinfo{booktitle}{{\em
  Proceedings of the 21st Annual International ACM SIGIR Conference on Research
  and Development in Information Retrieval}} {\em (\bibinfo{series}{SIGIR
  '98})}. \bibinfo{publisher}{ACM}, \bibinfo{address}{New York, NY, USA},
  \bibinfo{pages}{335--336}.
\newblock
\showISBNx{1-58113-015-5}
\showDOI{%
\url{https://doi.org/10.1145/290941.291025}}


\bibitem[\protect\citeauthoryear{Clemmensen and Borlund}{Clemmensen and
  Borlund}{2016}]%
        {borlund_order}
\bibfield{author}{\bibinfo{person}{Melanie~Landvad Clemmensen} {and}
  \bibinfo{person}{Pia Borlund}.} \bibinfo{year}{2016}\natexlab{}.
\newblock \showarticletitle{Order effect in interactive information retrieval
  evaluation: an empirical study}.
\newblock \bibinfo{journal}{{\em Journal of Documentation\/}}
  \bibinfo{volume}{72}, \bibinfo{number}{2} (\bibinfo{year}{2016}),
  \bibinfo{pages}{194--213}.
\newblock
\showDOI{%
\url{https://doi.org/10.1108/JD-04-2015-0051}}


\bibitem[\protect\citeauthoryear{Corporation}{Corporation}{1967}]%
        {cuadra_system1967experimental}
\bibfield{author}{\bibinfo{person}{System~Development Corporation}.}
  \bibinfo{year}{1967}\natexlab{}.
\newblock \bibinfo{booktitle}{{\em Experimental Studies of Relevance Judgments:
  Final Report}}.
\newblock \bibinfo{publisher}{System Development Corporation}.
\newblock
\showLCCN{78300675}
\showURL{%
\url{https://books.google.co.uk/books?id=80GRRwAACAAJ}}


\bibitem[\protect\citeauthoryear{Cosijn}{Cosijn}{2009}]%
        {cosijn2009relevance}
\bibfield{author}{\bibinfo{person}{Erica Cosijn}.}
  \bibinfo{year}{2009}\natexlab{}.
\newblock \showarticletitle{Relevance Judgments and Measurements}.
\newblock In \bibinfo{booktitle}{{\em Encyclopedia of Library and Information
  Sciences}}. \bibinfo{publisher}{CRC Press}, \bibinfo{pages}{4512--4519}.
\newblock


\bibitem[\protect\citeauthoryear{Cosijn and Ingwersen}{Cosijn and
  Ingwersen}{2000}]%
        {COSIJN2000533_dimensions}
\bibfield{author}{\bibinfo{person}{Erica Cosijn} {and} \bibinfo{person}{Peter
  Ingwersen}.} \bibinfo{year}{2000}\natexlab{}.
\newblock \showarticletitle{Dimensions of relevance}.
\newblock \bibinfo{journal}{{\em Information Processing \& Management\/}}
  \bibinfo{volume}{36}, \bibinfo{number}{4} (\bibinfo{year}{2000}),
  \bibinfo{pages}{533 -- 550}.
\newblock
\showISSN{0306-4573}
\showDOI{%
\url{https://doi.org/10.1016/S0306-4573(99)00072-2}}


\bibitem[\protect\citeauthoryear{da~Costa~Pereira, Dragoni, and
  Pasi}{da~Costa~Pereira et~al\mbox{.}}{2009}]%
        {pasi_new_aggre_crit}
\bibfield{author}{\bibinfo{person}{C{\'e}lia da Costa~Pereira},
  \bibinfo{person}{Mauro Dragoni}, {and} \bibinfo{person}{Gabriella Pasi}.}
  \bibinfo{year}{2009}\natexlab{}.
\newblock \showarticletitle{Multidimensional relevance: A new aggregation
  criterion}. In \bibinfo{booktitle}{{\em European Conference on Information
  Retrieval}}. Springer, \bibinfo{pages}{264--275}.
\newblock


\bibitem[\protect\citeauthoryear{da~Costa~Pereira, Dragoni, and
  Pasi}{da~Costa~Pereira et~al\mbox{.}}{2012}]%
        {daCostaPereira_pasi_priorit_agg_oper}
\bibfield{author}{\bibinfo{person}{C{\'e}lia da Costa~Pereira},
  \bibinfo{person}{Mauro Dragoni}, {and} \bibinfo{person}{Gabriella Pasi}.}
  \bibinfo{year}{2012}\natexlab{}.
\newblock \showarticletitle{Multidimensional Relevance: Prioritized Aggregation
  in a Personalized Information Retrieval Setting}.
\newblock \bibinfo{journal}{{\em Inf. Process. Manage.\/}}
  \bibinfo{volume}{48}, \bibinfo{number}{2} (\bibinfo{date}{March}
  \bibinfo{year}{2012}), \bibinfo{pages}{340--357}.
\newblock
\showISSN{0306-4573}
\showDOI{%
\url{https://doi.org/10.1016/j.ipm.2011.07.001}}


\bibitem[\protect\citeauthoryear{Eisenberg and Barry}{Eisenberg and
  Barry}{1988}]%
        {Eisenberg1988_order}
\bibfield{author}{\bibinfo{person}{Michael Eisenberg} {and}
  \bibinfo{person}{Carol Barry}.} \bibinfo{year}{1988}\natexlab{}.
\newblock \showarticletitle{Order effects: A study of the possible influence of
  presentation order on user judgments of document relevance}.
\newblock \bibinfo{journal}{{\em Journal of the American Society for
  Information Science\/}} \bibinfo{volume}{39}, \bibinfo{number}{5}
  (\bibinfo{date}{Sept.} \bibinfo{year}{1988}), \bibinfo{pages}{293--300}.
\newblock
\showDOI{%
\url{https://doi.org/10.1002/(sici)1097-4571(198809)39:5<293::aid-asi1>3.0.co;2-i}}


\bibitem[\protect\citeauthoryear{Fell, Dehdashti, Bruza, and Moreira}{Fell
  et~al\mbox{.}}{2019}]%
        {fell:dehdashti:bruza:moreira:2019}
\bibfield{author}{\bibinfo{person}{L. Fell}, \bibinfo{person}{S. Dehdashti},
  \bibinfo{person}{P.D. Bruza}, {and} \bibinfo{person}{C. Moreira}.}
  \bibinfo{year}{2019}\natexlab{}.
\newblock \showarticletitle{An Experimental Protocol to Derive and Validate a
  Quantum Model of Decision-Making}. In \bibinfo{booktitle}{{\em Proceedings of
  the 41st Annual Meeting of the Cognitive Science Society (COGSCI'19)}}.
\newblock


\bibitem[\protect\citeauthoryear{Fuhr, Giachanou, Grefenstette, Gurevych,
  Hanselowski, Jarvelin, Jones, Liu, Mothe, Nejdl, Peters, and Stein}{Fuhr
  et~al\mbox{.}}{2018}]%
        {Fuhr_info_nutrition}
\bibfield{author}{\bibinfo{person}{Norbert Fuhr}, \bibinfo{person}{Anastasia
  Giachanou}, \bibinfo{person}{Gregory Grefenstette}, \bibinfo{person}{Iryna
  Gurevych}, \bibinfo{person}{Andreas Hanselowski}, \bibinfo{person}{Kalervo
  Jarvelin}, \bibinfo{person}{Rosie Jones}, \bibinfo{person}{YiquN Liu},
  \bibinfo{person}{Josiane Mothe}, \bibinfo{person}{Wolfgang Nejdl},
  \bibinfo{person}{Isabella Peters}, {and} \bibinfo{person}{Benno Stein}.}
  \bibinfo{year}{2018}\natexlab{}.
\newblock \showarticletitle{An Information Nutritional Label for Online
  Documents}.
\newblock \bibinfo{journal}{{\em SIGIR Forum\/}} \bibinfo{volume}{51},
  \bibinfo{number}{3} (\bibinfo{date}{Feb.} \bibinfo{year}{2018}),
  \bibinfo{pages}{46--66}.
\newblock
\showISSN{0163-5840}
\showDOI{%
\url{https://doi.org/10.1145/3190580.3190588}}


\bibitem[\protect\citeauthoryear{Galv{\~{a}}o}{Galv{\~{a}}o}{2005}]%
        {Galvo2005_wigner}
\bibfield{author}{\bibinfo{person}{Ernesto~F. Galv{\~{a}}o}.}
  \bibinfo{year}{2005}\natexlab{}.
\newblock \showarticletitle{Discrete Wigner functions and quantum computational
  speedup}.
\newblock \bibinfo{journal}{{\em Physical Review A\/}} \bibinfo{volume}{71},
  \bibinfo{number}{4} (\bibinfo{date}{April} \bibinfo{year}{2005}).
\newblock
\showDOI{%
\url{https://doi.org/10.1103/physreva.71.042302}}


\bibitem[\protect\citeauthoryear{Hogarth and Einhorn}{Hogarth and
  Einhorn}{1992}]%
        {Hogarth1992}
\bibfield{author}{\bibinfo{person}{Robin~M Hogarth} {and}
  \bibinfo{person}{Hillel~J Einhorn}.} \bibinfo{year}{1992}\natexlab{}.
\newblock \showarticletitle{Order effects in belief updating: The
  belief-adjustment model}.
\newblock \bibinfo{journal}{{\em Cognitive Psychology\/}} \bibinfo{volume}{24},
  \bibinfo{number}{1} (\bibinfo{date}{jan} \bibinfo{year}{1992}),
  \bibinfo{pages}{1--55}.
\newblock
\showDOI{%
\url{https://doi.org/10.1016/0010-0285(92)90002-j}}


\bibitem[\protect\citeauthoryear{hsuan Huang and yu~Wang}{hsuan Huang and
  yu~Wang}{2004}]%
        {Huang2004_order}
\bibfield{author}{\bibinfo{person}{Mu hsuan Huang} {and} \bibinfo{person}{Hui
  yu Wang}.} \bibinfo{year}{2004}\natexlab{}.
\newblock \showarticletitle{The influence of document presentation order and
  number of documents judged on users{\textquotesingle} judgments of
  relevance}.
\newblock \bibinfo{journal}{{\em Journal of the American Society for
  Information Science and Technology\/}} \bibinfo{volume}{55},
  \bibinfo{number}{11} (\bibinfo{year}{2004}), \bibinfo{pages}{970--979}.
\newblock
\showDOI{%
\url{https://doi.org/10.1002/asi.20047}}


\bibitem[\protect\citeauthoryear{Jiang and Allan}{Jiang and Allan}{2016}]%
        {ja16_effort}
\bibfield{author}{\bibinfo{person}{J. Jiang} {and} \bibinfo{person}{J. Allan}.}
  \bibinfo{year}{2016}\natexlab{}.
\newblock \showarticletitle{Adaptive effort for search evaluation metrics}.
\newblock \bibinfo{journal}{{\em In ECIR '\/}}  \bibinfo{volume}{16}
  (\bibinfo{year}{2016}), \bibinfo{pages}{187--199}.
\newblock


\bibitem[\protect\citeauthoryear{Jiang, He, and Allan}{Jiang
  et~al\mbox{.}}{2017}]%
        {Jiang2017_in_situ}
\bibfield{author}{\bibinfo{person}{Jiepu Jiang}, \bibinfo{person}{Daqing He},
  {and} \bibinfo{person}{James Allan}.} \bibinfo{year}{2017}\natexlab{}.
\newblock \showarticletitle{Comparing In Situ and Multidimensional Relevance
  Judgments}. In \bibinfo{booktitle}{{\em Proceedings of the 40th International
  {ACM} {SIGIR} Conference on Research and Development in Information Retrieval
  - {SIGIR} {\textquotesingle}17}}. \bibinfo{publisher}{{ACM} Press}.
\newblock
\showDOI{%
\url{https://doi.org/10.1145/3077136.3080840}}


\bibitem[\protect\citeauthoryear{Li, Zhang, Song, and Wu}{Li
  et~al\mbox{.}}{2017}]%
        {ASI:Jingfei}
\bibfield{author}{\bibinfo{person}{Jingfei Li}, \bibinfo{person}{Peng Zhang},
  \bibinfo{person}{Dawei Song}, {and} \bibinfo{person}{Yue Wu}.}
  \bibinfo{year}{2017}\natexlab{}.
\newblock \showarticletitle{Understanding an enriched multidimensional user
  relevance model by analyzing query logs}.
\newblock \bibinfo{journal}{{\em Journal of the Association for Information
  Science and Technology\/}} \bibinfo{volume}{68}, \bibinfo{number}{12}
  (\bibinfo{year}{2017}), \bibinfo{pages}{2743--2754}.
\newblock
\showISSN{2330-1643}
\showDOI{%
\url{https://doi.org/10.1002/asi.23868}}


\bibitem[\protect\citeauthoryear{Nilan and Fletcher}{Nilan and
  Fletcher}{1987}]%
        {nilan_info_behav}
\bibfield{author}{\bibinfo{person}{{Michael S.} Nilan} {and}
  \bibinfo{person}{{Patricia T.} Fletcher}.} \bibinfo{year}{1987}\natexlab{}.
\newblock \showarticletitle{INFORMATION BEHAVIORS IN THE PREPARATION OF
  RESEARCH PROPOSALS: A USER STUDY.}. In \bibinfo{booktitle}{{\em Proceedings
  of the ASIS Annual Meeting}}, \bibfield{editor}{\bibinfo{person}{Ching chih
  Chen}} (Ed.), Vol.~\bibinfo{volume}{24}. \bibinfo{publisher}{Learned
  Information Inc}, \bibinfo{pages}{186--192}.
\newblock
\showISBNx{0938734199}


\bibitem[\protect\citeauthoryear{Nilan, Peek, and Snyder}{Nilan
  et~al\mbox{.}}{1988}]%
        {nilan1988methodology}
\bibfield{author}{\bibinfo{person}{M.~S. Nilan}, \bibinfo{person}{R.~P. Peek},
  {and} \bibinfo{person}{H.~W. Snyder}.} \bibinfo{year}{1988}\natexlab{}.
\newblock \showarticletitle{A Methodology for Tapping User Evaluation
  Behaviors: An Exploration of Users\textquotesingle Strategy, Source and
  Information Evaluating}.
\newblock In \bibinfo{booktitle}{{\em Proceedings of the American Society for
  Information Science ({ASIS}) 51st Annual Meeting}},
  \bibfield{editor}{\bibinfo{person}{C.~L. Borgman} {and}
  \bibinfo{person}{E.~Y.~H. Pai}} (Eds.). \bibinfo{publisher}{Learned
  Information}, \bibinfo{address}{Medford, NJ}, \bibinfo{pages}{152--159}.
\newblock


\bibitem[\protect\citeauthoryear{Olteanu, Peshterliev, Liu, and Aberer}{Olteanu
  et~al\mbox{.}}{2013}]%
        {opla13_rel}
\bibfield{author}{\bibinfo{person}{A. Olteanu}, \bibinfo{person}{S.
  Peshterliev}, \bibinfo{person}{X. Liu}, {and} \bibinfo{person}{K. Aberer}.}
  \bibinfo{year}{2013}\natexlab{}.
\newblock \showarticletitle{Web credibility: Features exploration and
  credibility prediction}.
\newblock \bibinfo{journal}{{\em In ECIR '\/}}  \bibinfo{volume}{13}
  (\bibinfo{year}{2013}), \bibinfo{pages}{557--568}.
\newblock


\bibitem[\protect\citeauthoryear{Palotti, Goeuriot, Zuccon, and
  Hanbury}{Palotti et~al\mbox{.}}{2016}]%
        {Palotti_understandability}
\bibfield{author}{\bibinfo{person}{Joao Palotti}, \bibinfo{person}{Lorraine
  Goeuriot}, \bibinfo{person}{Guido Zuccon}, {and} \bibinfo{person}{Allan
  Hanbury}.} \bibinfo{year}{2016}\natexlab{}.
\newblock \showarticletitle{Ranking Health Web Pages with Relevance and
  Understandability}. In \bibinfo{booktitle}{{\em Proceedings of the 39th
  International ACM SIGIR Conference on Research and Development in Information
  Retrieval}} {\em (\bibinfo{series}{SIGIR '16})}. \bibinfo{publisher}{ACM},
  \bibinfo{address}{New York, NY, USA}, \bibinfo{pages}{965--968}.
\newblock
\showISBNx{978-1-4503-4069-4}
\showDOI{%
\url{https://doi.org/10.1145/2911451.2914741}}


\bibitem[\protect\citeauthoryear{Park}{Park}{1993}]%
        {park1993nature_of_rel}
\bibfield{author}{\bibinfo{person}{Taemin~Kim Park}.}
  \bibinfo{year}{1993}\natexlab{}.
\newblock \showarticletitle{The nature of relevance in information retrieval:
  An empirical study}.
\newblock \bibinfo{journal}{{\em The library quarterly\/}}
  \bibinfo{volume}{63}, \bibinfo{number}{3} (\bibinfo{year}{1993}),
  \bibinfo{pages}{318--351}.
\newblock


\bibitem[\protect\citeauthoryear{Pothos and Busemeyer}{Pothos and
  Busemeyer}{2009}]%
        {Pothos2009_quan_explan_irrational}
\bibfield{author}{\bibinfo{person}{E.~M. Pothos} {and} \bibinfo{person}{J.~R.
  Busemeyer}.} \bibinfo{year}{2009}\natexlab{}.
\newblock \showarticletitle{A quantum probability explanation for violations of
  {\textquotesingle}rational{\textquotesingle} decision theory}.
\newblock \bibinfo{journal}{{\em Proceedings of the Royal Society B: Biological
  Sciences\/}} \bibinfo{volume}{276}, \bibinfo{number}{1665}
  (\bibinfo{date}{mar} \bibinfo{year}{2009}), \bibinfo{pages}{2171--2178}.
\newblock
\showDOI{%
\url{https://doi.org/10.1098/rspb.2009.0121}}


\bibitem[\protect\citeauthoryear{Sakurai and Napolitano}{Sakurai and
  Napolitano}{2017}]%
        {sakurai}
\bibfield{author}{\bibinfo{person}{J.~J. Sakurai} {and} \bibinfo{person}{Jim
  Napolitano}.} \bibinfo{year}{2017}\natexlab{}.
\newblock \bibinfo{booktitle}{{\em Modern Quantum Mechanics}}.
\newblock \bibinfo{publisher}{Cambridge University Press}.
\newblock
\showISBNx{1108422411}


\bibitem[\protect\citeauthoryear{Saracevic}{Saracevic}{1996}]%
        {saracevic1996_stratified}
\bibfield{author}{\bibinfo{person}{Tefko Saracevic}.}
  \bibinfo{year}{1996}\natexlab{}.
\newblock \showarticletitle{Modeling Interaction in Information Retrieval (IR):
  A Review and Proposal.}. In \bibinfo{booktitle}{{\em Proceedings of the ASIS
  annual meeting}}, Vol.~\bibinfo{volume}{33}. ERIC, \bibinfo{pages}{3--9}.
\newblock


\bibitem[\protect\citeauthoryear{Saracevic}{Saracevic}{1997}]%
        {saracevic1997stratified}
\bibfield{author}{\bibinfo{person}{Tefko Saracevic}.}
  \bibinfo{year}{1997}\natexlab{}.
\newblock \showarticletitle{The stratified model of information retrieval
  interaction: Extension and applications}.
\newblock


\bibitem[\protect\citeauthoryear{Saracevic}{Saracevic}{2007}]%
        {Saracevic2007_part2}
\bibfield{author}{\bibinfo{person}{Tefko Saracevic}.}
  \bibinfo{year}{2007}\natexlab{}.
\newblock \showarticletitle{Relevance: A review of the literature and a
  framework for thinking on the notion in information science. Part {II}:
  nature and manifestations of relevance}.
\newblock \bibinfo{journal}{{\em Journal of the American Society for
  Information Science and Technology\/}} \bibinfo{volume}{58},
  \bibinfo{number}{13} (\bibinfo{year}{2007}), \bibinfo{pages}{1915--1933}.
\newblock
\showDOI{%
\url{https://doi.org/10.1002/asi.20682}}


\bibitem[\protect\citeauthoryear{Saracevic}{Saracevic}{2016}]%
        {Saracevic2016_notion}
\bibfield{author}{\bibinfo{person}{Tefko Saracevic}.}
  \bibinfo{year}{2016}\natexlab{}.
\newblock \showarticletitle{The Notion of Relevance in Information Science:
  Everybody knows what relevance is. But, what is it really?}
\newblock \bibinfo{journal}{{\em Synthesis Lectures on Information Concepts,
  Retrieval, and Services\/}} \bibinfo{volume}{8}, \bibinfo{number}{3}
  (\bibinfo{date}{Sept.} \bibinfo{year}{2016}), \bibinfo{pages}{i--109}.
\newblock
\showDOI{%
\url{https://doi.org/10.2200/s00723ed1v01y201607icr050}}


\bibitem[\protect\citeauthoryear{Schwarz and Morris}{Schwarz and
  Morris}{2011}]%
        {sm11_rel}
\bibfield{author}{\bibinfo{person}{J. Schwarz} {and} \bibinfo{person}{M.
  Morris}.} \bibinfo{year}{2011}\natexlab{}.
\newblock \showarticletitle{Augmenting web pages and search results to support
  credibility assessment}.
\newblock \bibinfo{journal}{{\em In CHI '\/}}  \bibinfo{volume}{11}
  (\bibinfo{year}{2011}), \bibinfo{pages}{1245--1254}.
\newblock


\bibitem[\protect\citeauthoryear{Slovic}{Slovic}{1995}]%
        {Slovic1995_construc_pref}
\bibfield{author}{\bibinfo{person}{Paul Slovic}.}
  \bibinfo{year}{1995}\natexlab{}.
\newblock \showarticletitle{The construction of preference.}
\newblock \bibinfo{journal}{{\em American Psychologist\/}}
  \bibinfo{volume}{50}, \bibinfo{number}{5} (\bibinfo{year}{1995}),
  \bibinfo{pages}{364--371}.
\newblock
\showDOI{%
\url{https://doi.org/10.1037/0003-066x.50.5.364}}


\bibitem[\protect\citeauthoryear{Trueblood and Busemeyer}{Trueblood and
  Busemeyer}{2011}]%
        {Trueblood2011_quantum_account_ordereff}
\bibfield{author}{\bibinfo{person}{Jennifer~S. Trueblood} {and}
  \bibinfo{person}{Jerome~R. Busemeyer}.} \bibinfo{year}{2011}\natexlab{}.
\newblock \showarticletitle{A Quantum Probability Account of Order Effects in
  Inference}.
\newblock \bibinfo{journal}{{\em Cognitive Science\/}} \bibinfo{volume}{35},
  \bibinfo{number}{8} (\bibinfo{date}{sep} \bibinfo{year}{2011}),
  \bibinfo{pages}{1518--1552}.
\newblock
\showDOI{%
\url{https://doi.org/10.1111/j.1551-6709.2011.01197.x}}


\bibitem[\protect\citeauthoryear{Tversky and Kahneman}{Tversky and
  Kahneman}{1974}]%
        {Tversky1974}
\bibfield{author}{\bibinfo{person}{A. Tversky} {and} \bibinfo{person}{D.
  Kahneman}.} \bibinfo{year}{1974}\natexlab{}.
\newblock \showarticletitle{Judgment under Uncertainty: Heuristics and Biases}.
\newblock \bibinfo{journal}{{\em Science\/}} \bibinfo{volume}{185},
  \bibinfo{number}{4157} (\bibinfo{date}{Sept.} \bibinfo{year}{1974}),
  \bibinfo{pages}{1124--1131}.
\newblock
\showDOI{%
\url{https://doi.org/10.1126/science.185.4157.1124}}


\bibitem[\protect\citeauthoryear{Uprety and Song}{Uprety and Song}{2018}]%
        {Uprety:2018:IOE:3234944.3234972}
\bibfield{author}{\bibinfo{person}{Sagar Uprety} {and} \bibinfo{person}{Dawei
  Song}.} \bibinfo{year}{2018}\natexlab{}.
\newblock \showarticletitle{Investigating Order Effects in Multidimensional
  Relevance Judgment Using Query Logs}. In \bibinfo{booktitle}{{\em Proceedings
  of the 2018 ACM SIGIR International Conference on Theory of Information
  Retrieval}} {\em (\bibinfo{series}{ICTIR '18})}. \bibinfo{publisher}{ACM},
  \bibinfo{address}{New York, NY, USA}, \bibinfo{pages}{191--194}.
\newblock
\showISBNx{978-1-4503-5656-5}
\showDOI{%
\url{https://doi.org/10.1145/3234944.3234972}}


\bibitem[\protect\citeauthoryear{Uprety, Su, Song, and Li}{Uprety
  et~al\mbox{.}}{2018}]%
        {Uprety:2018:MMU:3209978.3210130}
\bibfield{author}{\bibinfo{person}{Sagar Uprety}, \bibinfo{person}{Yi Su},
  \bibinfo{person}{Dawei Song}, {and} \bibinfo{person}{Jingfei Li}.}
  \bibinfo{year}{2018}\natexlab{}.
\newblock \showarticletitle{Modeling Multidimensional User Relevance in IR
  Using Vector Spaces}. In \bibinfo{booktitle}{{\em The 41st International ACM
  SIGIR Conference on Research \& Development in Information Retrieval}} {\em
  (\bibinfo{series}{SIGIR '18})}. \bibinfo{publisher}{ACM},
  \bibinfo{address}{New York, NY, USA}, \bibinfo{pages}{993--996}.
\newblock
\showISBNx{978-1-4503-5657-2}
\showDOI{%
\url{https://doi.org/10.1145/3209978.3210130}}


\bibitem[\protect\citeauthoryear{Verma, Yilmaz, and Craswell}{Verma
  et~al\mbox{.}}{2016}]%
        {vyc16_effort}
\bibfield{author}{\bibinfo{person}{M. Verma}, \bibinfo{person}{E. Yilmaz},
  {and} \bibinfo{person}{N. Craswell}.} \bibinfo{year}{2016}\natexlab{}.
\newblock \showarticletitle{On obtaining effort based judgements for
  information retrieval}.
\newblock \bibinfo{journal}{{\em In WSDM '\/}}  \bibinfo{volume}{16}
  (\bibinfo{year}{2016}), \bibinfo{pages}{277--286}.
\newblock


\bibitem[\protect\citeauthoryear{Wang, Zhang, Li, Song, Hou, and Shang}{Wang
  et~al\mbox{.}}{2016}]%
        {benyou_quantum_interf_Order}
\bibfield{author}{\bibinfo{person}{Benyou Wang}, \bibinfo{person}{Peng Zhang},
  \bibinfo{person}{Jingfei Li}, \bibinfo{person}{Dawei Song},
  \bibinfo{person}{Yuexian Hou}, {and} \bibinfo{person}{Zhenguo Shang}.}
  \bibinfo{year}{2016}\natexlab{}.
\newblock \showarticletitle{Exploration of Quantum Interference in Document
  Relevance Judgement Discrepancy}.
\newblock \bibinfo{journal}{{\em Entropy\/}} \bibinfo{volume}{18},
  \bibinfo{number}{12} (\bibinfo{date}{Apr} \bibinfo{year}{2016}),
  \bibinfo{pages}{144}.
\newblock
\showISSN{1099-4300}
\showDOI{%
\url{https://doi.org/10.3390/e18040144}}


\bibitem[\protect\citeauthoryear{Wang and Busemeyer}{Wang and
  Busemeyer}{2013}]%
        {Wang2013}
\bibfield{author}{\bibinfo{person}{Zheng Wang} {and} \bibinfo{person}{Jerome~R.
  Busemeyer}.} \bibinfo{year}{2013}\natexlab{}.
\newblock \showarticletitle{A Quantum Question Order Model Supported by
  Empirical Tests of {anA} Prioriand Precise Prediction}.
\newblock \bibinfo{journal}{{\em Topics in Cognitive Science\/}}
  (\bibinfo{date}{sep} \bibinfo{year}{2013}).
\newblock
\showDOI{%
\url{https://doi.org/10.1111/tops.12040}}


\bibitem[\protect\citeauthoryear{Wawer, Nielek, and Wierzbicki}{Wawer
  et~al\mbox{.}}{2014}]%
        {wnw14_rel}
\bibfield{author}{\bibinfo{person}{A. Wawer}, \bibinfo{person}{R. Nielek},
  {and} \bibinfo{person}{A. Wierzbicki}.} \bibinfo{year}{2014}\natexlab{}.
\newblock \showarticletitle{Predicting webpage credibility using linguistic
  features}.
\newblock \bibinfo{journal}{{\em In WWW '\/}}  \bibinfo{volume}{14}
  (\bibinfo{year}{2014}), \bibinfo{pages}{1135--1140}.
\newblock


\bibitem[\protect\citeauthoryear{Wootters}{Wootters}{1987}]%
        {Wootters1987_wigner}
\bibfield{author}{\bibinfo{person}{William~K Wootters}.}
  \bibinfo{year}{1987}\natexlab{}.
\newblock \showarticletitle{A Wigner-function formulation of finite-state
  quantum mechanics}.
\newblock \bibinfo{journal}{{\em Annals of Physics\/}} \bibinfo{volume}{176},
  \bibinfo{number}{1} (\bibinfo{date}{May} \bibinfo{year}{1987}),
  \bibinfo{pages}{1--21}.
\newblock
\showDOI{%
\url{https://doi.org/10.1016/0003-4916(87)90176-x}}


\bibitem[\protect\citeauthoryear{Xu and Wang}{Xu and Wang}{2008}]%
        {Xu2008_order}
\bibfield{author}{\bibinfo{person}{Yunjie Xu} {and} \bibinfo{person}{Dong
  Wang}.} \bibinfo{year}{2008}\natexlab{}.
\newblock \showarticletitle{Order effect in relevance judgment}.
\newblock \bibinfo{journal}{{\em Journal of the American Society for
  Information Science and Technology\/}} \bibinfo{volume}{59},
  \bibinfo{number}{8} (\bibinfo{year}{2008}), \bibinfo{pages}{1264--1275}.
\newblock
\showDOI{%
\url{https://doi.org/10.1002/asi.20826}}


\bibitem[\protect\citeauthoryear{Xu and Chen}{Xu and Chen}{2006}]%
        {ASI:Xu-MURM}
\bibfield{author}{\bibinfo{person}{Yunjie~(Calvin) Xu} {and}
  \bibinfo{person}{Zhiwei Chen}.} \bibinfo{year}{2006}\natexlab{}.
\newblock \showarticletitle{Relevance judgment: What do information users
  consider beyond topicality?}
\newblock \bibinfo{journal}{{\em Journal of the American Society for
  Information Science and Technology\/}} \bibinfo{volume}{57},
  \bibinfo{number}{7} (\bibinfo{year}{2006}), \bibinfo{pages}{961--973}.
\newblock
\showISSN{1532-2890}
\showDOI{%
\url{https://doi.org/10.1002/asi.20361}}


\bibitem[\protect\citeauthoryear{Yamamoto and Tanaka}{Yamamoto and
  Tanaka}{2011}]%
        {yt11_rel}
\bibfield{author}{\bibinfo{person}{Y. Yamamoto} {and} \bibinfo{person}{K.
  Tanaka}.} \bibinfo{year}{2011}\natexlab{}.
\newblock \showarticletitle{Enhancing credibility judgment of web search
  results}.
\newblock \bibinfo{journal}{{\em In CHI '\/}}  \bibinfo{volume}{11}
  (\bibinfo{year}{2011}), \bibinfo{pages}{1235--1244}.
\newblock


\bibitem[\protect\citeauthoryear{Yilmaz, Verma, Craswell, Radlinski, and
  Bailey}{Yilmaz et~al\mbox{.}}{2014}]%
        {yvcrb14_effort}
\bibfield{author}{\bibinfo{person}{E. Yilmaz}, \bibinfo{person}{M. Verma},
  \bibinfo{person}{N. Craswell}, \bibinfo{person}{F. Radlinski}, {and}
  \bibinfo{person}{P. Bailey}.} \bibinfo{year}{2014}\natexlab{}.
\newblock \showarticletitle{Relevance and effort: An analysis of document
  utility}.
\newblock \bibinfo{journal}{{\em In CIKM '\/}}  \bibinfo{volume}{14}
  (\bibinfo{year}{2014}), \bibinfo{pages}{91--100}.
\newblock


\bibitem[\protect\citeauthoryear{Zhai, Cohen, and Lafferty}{Zhai
  et~al\mbox{.}}{2003}]%
        {zhai2003beyond_diversity}
\bibfield{author}{\bibinfo{person}{Cheng~Xiang Zhai},
  \bibinfo{person}{William~W Cohen}, {and} \bibinfo{person}{John Lafferty}.}
  \bibinfo{year}{2003}\natexlab{}.
\newblock \showarticletitle{Beyond independent relevance: methods and
  evaluation metrics for subtopic retrieval}. In \bibinfo{booktitle}{{\em
  Proceedings of the 26th annual international ACM SIGIR conference on Research
  and development in informaion retrieval}}. ACM, \bibinfo{pages}{10--17}.
\newblock


\bibitem[\protect\citeauthoryear{Zhang, Zhang, Lease, and Gwizdka}{Zhang
  et~al\mbox{.}}{2014}]%
        {MURM-psychometrics}
\bibfield{author}{\bibinfo{person}{Yinglong Zhang}, \bibinfo{person}{Jin
  Zhang}, \bibinfo{person}{Matthew Lease}, {and} \bibinfo{person}{Jacek
  Gwizdka}.} \bibinfo{year}{2014}\natexlab{}.
\newblock \showarticletitle{Multidimensional Relevance Modeling via
  Psychometrics and Crowdsourcing}. In \bibinfo{booktitle}{{\em Proceedings of
  the 37th International ACM SIGIR Conference on Research \& Development in
  Information Retrieval}} {\em (\bibinfo{series}{SIGIR '14})}.
  \bibinfo{publisher}{ACM}, \bibinfo{address}{New York, NY, USA},
  \bibinfo{pages}{435--444}.
\newblock
\showISBNx{978-1-4503-2257-7}
\showDOI{%
\url{https://doi.org/10.1145/2600428.2609577}}


\bibitem[\protect\citeauthoryear{Zuccon}{Zuccon}{2016}]%
        {Understandability_Guido}
\bibfield{author}{\bibinfo{person}{Guido Zuccon}.}
  \bibinfo{year}{2016}\natexlab{}.
\newblock \showarticletitle{Understandability Biased Evaluation for Information
  Retrieval}. In \bibinfo{booktitle}{{\em Advances in Information Retrieval}},
  \bibfield{editor}{\bibinfo{person}{Nicola Ferro}, \bibinfo{person}{Fabio
  Crestani}, \bibinfo{person}{Marie-Francine Moens}, \bibinfo{person}{Josiane
  Mothe}, \bibinfo{person}{Fabrizio Silvestri}, \bibinfo{person}{Giorgio~Maria
  Di~Nunzio}, \bibinfo{person}{Claudia Hauff}, {and} \bibinfo{person}{Gianmaria
  Silvello}} (Eds.). \bibinfo{publisher}{Springer International Publishing},
  \bibinfo{address}{Cham}, \bibinfo{pages}{280--292}.
\newblock


\end{thebibliography}
